\documentclass[twocolumn,english,aps,prl,showpacs,floatfix,superscriptaddress]{revtex4-1}
\usepackage[T1]{fontenc}
\usepackage[latin9]{inputenc}
\usepackage{amsmath}

\makeatletter
\usepackage{microtype}

\usepackage{dcolumn}
\usepackage[english]{babel}
\usepackage{graphicx}
\usepackage[small,bf,hang,raggedright,FIGTOPCAP]{subfigure}
\usepackage{varioref}
\usepackage{epstopdf}
\usepackage{color}
\usepackage{eucal}
\usepackage{float}
\usepackage{soul}
\usepackage{physics}
\graphicspath{{./pictures/}}

\makeatother

\usepackage{babel}
\begin{document}

\title{Transport Waves as Crystal Excitations}

\date{\today}

\author{Andrea Cepellotti}
\affiliation{Theory and Simulations of Materials (THEOS) and National Centre for Computational Design and Discovery of Novel Materials
	(MARVEL), \'Ecole Polytechnique
	F\'ed\'erale de Lausanne, Station 9, 1015 Lausanne, Switzerland}
\affiliation{Department of Physics, University of California at Berkeley, Berkeley, California 94720, United States}
\affiliation{Materials Science Division, Lawrence Berkeley National Laboratory, Berkeley, California 94720, United States}
\author{Nicola Marzari}
\affiliation{Theory and Simulations of Materials (THEOS) and National Centre for Computational Design and Discovery of Novel Materials
(MARVEL), \'Ecole Polytechnique
F\'ed\'erale de Lausanne, Station 9, 1015 Lausanne, Switzerland}

\begin{abstract}
We introduce the concept of transport waves by showing that the linearized Boltzmann transport equation admits excitations in the form of waves that have well defined dispersion relations and decay times. 
Crucially, these waves do not represent single-particle excitations, but are collective excitations of the equilibrium distribution functions.
We study in detail the case of thermal transport, where relaxons are found in the long-wavelength limit, and second sound is reinterpreted as the excitation of one or several temperature waves at finite frequencies. 
Graphene is studied numerically, finding decay times of the order of microseconds. 
The derivation, obtained by a spectral representation of the Boltzmann equation, holds in principle for any crystal or semiclassical transport theory and is particularly relevant when transport takes place in the hydrodynamic regime.
\end{abstract}
\maketitle

\section{Introduction}

Classical theories of transport are based on the Boltzmann transport equation (BTE), that abstracts the microscopic
dynamics of the carriers in an out-of-equilibrium system into a dynamics of the carriers' distributions.
In its complete form it is a non-linear integro-differential equation that includes a drift-diffusion and
a scattering term; close to equilibrium it simplifies in the linearized Boltzmann transport equation (LBTE), with the scattering
term taking a linear matricial form \cite{ziman}. As such, it has often been applied in a semiclassical form to
study transport phenomena in crystals, where the carriers are e.g. phonon or electron wavepackets with
well defined energies and quasimomenta, and where the scattering rates or the lifetimes can be calculated from
model or first-principles interactions. For the case of thermal transport, in particular, a number of fairly exotic
phenomena have been studied both experimentally and theoretically \cite{cahill:review}. One of these is the propagation of 
second sound: a temperature wave that has been observed in a handful of materials at cryogenic conditions, namely solid helium  \cite{ss:he}, sodium fluoride \cite{ss:naf1,ss:naf2}, bismuth \cite{ss:bi}, sapphire \cite{sapphire:ss}, and strontium titanate \cite{ss:srtio1,ss:srtio2}, and has been suggested to
exist in two-dimensional or layered materials \cite{fugallo-nano,nature-chen,nature-mio}.

Despite the fact that second sound has been known for decades, its microscopic underpinnings have proven challenging, while
evoking some paradigmatic, and fundamental, collective behaviour. In fact,
second sound arises out of a collective response of the phonon `gas', so that the challenge arises in using microscopic degrees of freedom to build a macroscopic equation for a damped temperature wave.
The earliest attempts \cite{chester,ss:kwok,ss:griffin,ss:enz} relied on ad-hoc assumptions, namely the introduction of some inertial term as an intrinsic property of the phonon gas.
Subsequent studies \cite{krumhansl:ss,guyer:ss,guyer:pr1} have shown that the phonon LBTE can be used as the starting point to derive a damped temperature wave equation.
In particular, it was shown that a material can host second sound when normal (momentum conserving) scattering events are much more frequent than Umklapp processes (where a quantity of momentum equal to a reciprocal lattice vector can be lost); this can happen at low temperatures or in 2D materials.
Few years later, it was shown \cite{hardy-ss} that this condition on normal and Umklapp processes isn't necessary for the existence of second sound; more generally, second sound relies on the existence of a mechanism by which the energy flux decays slowly enough so that the crystal can sustain a temperature wave for long times.
%However, existing studies rely on assumptions or approximations of the LBTE and thus the precise nature of these collective excitations is still left unclear.
We refer the reader to Ref. \cite{heatwaves:revmodphys} for a thorough review of studies until 1989.
We note however that all existing studies of second sound rely on simplifications of the LBTE: in most cases the description of phonon scattering is simplified, for example making use of the relaxation time approximation or the Callaway approximation \cite{lee:secondsoundnew}. 
To our knowledge, only Hardy \cite{hardy-ss} attempted a study of second sound using the complete LBTE, but resorted to assumptions on the eigenvalue spectrum of the scattering matrix and limited his study to systems of cubic symmetry.
These simplifications are not necessary and it is our aim to characterize second sound by solving exactly the LBTE.

In this work, we show very generally that the LBTE admits the existence of excitations in the form of propagating waves for the carriers distributions.
%By means of a spectral representation, we recast the solution of the LBTE in the form of an eigenvalue equation, where each damped oscillator identifies a collective excitation (the transport wave) with well defined dispersion relations and decay times. 
By means of a spectral representation, we recast the solution of the LBTE in the form of an eigenvalue equation, where each damped oscillator identifies an excitation (the transport wave) with well defined dispersion relations and decay times. 
We stress that these waves do not represent single particle excitations, but are collective excitations of the equilibrium distribution functions. % {\color{red}Crucially, such damped oscillators are not single-phonon oscillators, but collective phonon excitations.} 
For the case of thermal transport, these excitations represent energy (heat) or temperature waves, that in the long wavelength limit reduce to relaxons \cite{relaxons}, i.e. the heat carriers of bulk steady state transport.
The proof does not rely on any particular assumption in the LBTE, so that temperature waves exist in principle in every dielectric crystal, if for a short time, and analogue transport excitations would be present in other different models of semiclassical transport.

%%%%%%%%%%%%%%%%%%%%%%%%%%%%%%%%%%%%%%%%%%%%%%%%%%%%%%%%%%%%%%%%%%%%%%%%%%%%%%%%%%%%%%

\section{Transport waves and second sound}

To begin with, we focus on thermal transport, and recall that the state of crystal lattice vibrations is determined by the phonon excitation number $n_{\mu}(\boldsymbol{x},t)$ for any possible phonon state $\mu$ (a shorthand notation to label $\mu=(\boldsymbol{q},s)$ with $\boldsymbol{q}$ the phonon wavevector and $s$ the phonon branch) at position $\boldsymbol{x}$ and time $t$.
At equilibrium, the phonon excitation number is given by the Bose--Einstein distribution function $\bar{n}_{\mu}=\frac{1}{\exp(\hbar\omega_{\mu}/k_{B}T_{0})-1}$, with $T_{0}$ the temperature of the crystal and $\omega_{\mu}$ the phonon frequency.
Out of thermal equilibrium, one observes a deviation of the phonon excitation numbers $\Delta n_{\mu}(\boldsymbol{x},t)=n_{\mu}(\boldsymbol{x},t)-\bar{n}_{\mu}$.
Note that, keeping the crystal at constant temperature, the Bose--Einstein distribution does not depend on space or time.

Temperature waves are directly related to oscillations in the phonon excitation numbers.
To see this, note that $\Delta n_\mu$ induces a change $\Delta E$ to the total energy $E$ of the crystal (i.e. heat):
\begin{align}
E(\boldsymbol{x},t) & =%\sum_{\mu}\Big(n_{\mu}(\boldsymbol{x},t)+\frac{1}{2}\Big)\hbar\omega_{\mu}\nonumber \\
 \sum_{\mu}\Big(\bar{n}_{\mu}+\frac{1}{2}\Big)\hbar\omega_{\mu}+\sum_{\mu}\Delta n_{\mu}(\boldsymbol{x},t)\hbar\omega_{\mu}\nonumber \\
 & =E_{0}+\Delta E(\boldsymbol{x},t)\;,
\end{align}
where $E_{0}$ is the energy at thermal equilibrium.
Therefore, oscillations in $\Delta n_\mu$ are carried over to $\Delta E$.
From kinetic arguments \cite{hardy-ss}, a change in energy corresponds to a change in temperature through the relation $\Delta E=C\Delta T$, where $C$ is the specific heat.
Therefore, to find an equation for a temperature wave, one must look for waves in the phonon excitation numbers $\Delta n_{\mu}$.

One can use the LBTE to describe the dynamics of $\Delta n_{\mu}$ \cite{relaxons}:
\begin{equation}
\frac{\partial\Delta n_{\mu}(\boldsymbol{x},t)}{\partial t}+\boldsymbol{v}_{\mu}\cdot\nabla(\Delta n_{\mu}(\boldsymbol{x},t))=-\frac{1}{\mathcal{V}}\sum_{\mu'}\Omega_{\mu\mu'}\Delta n_{\mu'}(\boldsymbol{x},t)\;,
\label{phonon_bte}
\end{equation}
where $\mathcal{V}$ is a normalization volume, $\boldsymbol{v}_{\mu}$ the phonon group velocity and $\Omega_{\mu\mu'}$ the scattering matrix (one can derive (2) from Eq. 2 of Ref. \cite{relaxons} by setting temperature derivatives to zero, since the average crystal temperature $T_0$ is constant).
The scattering operator contains the rates for all transitions $\mu\to\mu'$ and will be later constructed with first-principles scattering rates of three-phonon and phonon-isotope processes (for their expressions, we refer to Refs. \cite{fugalloPRB,relaxons,marzari:prl}).

We now look for wave solutions for the phonon excitation numbers: 
\begin{equation}
\Delta n_{\mu}=\Re\Big(I_{\mu}e^{i(\boldsymbol{k}\cdot\boldsymbol{r}-\omega t)}\Big)\;,
\end{equation}
where $I_{\mu}$ is the wave amplitude for the mode $\mu$, $\boldsymbol{k}$ is the wavevector and $\omega$ is the frequency (to not confuse with the phonon frequency $\omega_{\mu}$).
The LBTE is most conveniently solved in the complex plane and then projected on the real axis ($\Re$ denotes the real part).
Inserting this Ansatz in Eq. \ref{phonon_bte}, i.e. applying a Fourier transform, one finds:
\begin{equation}
-i\omega I_{\mu}+iI_{\mu}\boldsymbol{k}\cdot\boldsymbol{v}_{\mu}+\frac{1}{\mathcal{V}}\sum_{\mu'}\Omega_{\mu\mu'}I_{\mu'}=0\;.
\label{eq4}\end{equation}
Rearranging the terms, Eq. \ref{eq4} can be recast as an eigenvalue problem: 
\begin{equation}
\sum_{\mu'}B_{\mu\mu'}(\boldsymbol{k})I_{\mu'}^{\alpha}(\boldsymbol{k})=\omega_{\alpha}(\boldsymbol{k})I_{\mu}^{\alpha}(\boldsymbol{k})\;, \label{eigenvalue}
\end{equation}
where
\begin{equation}
B_{\mu\mu'}=\boldsymbol{k}\cdot\boldsymbol{v}_{\mu}\delta_{\mu\mu'}-\frac{i}{\mathcal{V}}\Omega_{\mu\mu'}\;,
\end{equation}
and $\alpha$ is the eigenvalue index.
This equation represents the central result and shows that the LBTE admits temperature-wave solutions whenever $I_{\mu}^{\alpha}(\boldsymbol{k})$ is a right eigenvector of $B(\boldsymbol{k})$ with $\omega_{\alpha}(\boldsymbol{k})$ as a corresponding eigenvalue.
Eq. \ref{eigenvalue} identifies with its eigenvectors a set of crystal excitations which appear in the form of oscillators characterized by a dispersion relation.

The matrix $B$ is complex, non-Hermitian and can be written  in a symmetric form by means of a simple scaling of variables (the same transformation used in Ref. \cite{relaxons}); in order to keep the main text as simple as possible we only discuss the transformation in the Appendix, although later numerical results rely on such symmetrized form.
As a consequence of these properties, eigenvalues $\omega_{\alpha}(\boldsymbol{k})$ are complex.
Therefore, it's more convenient to write them as
\begin{equation}
\omega_{\alpha}(\boldsymbol{k})=\overline{\omega}_{\alpha}(\boldsymbol{k})-\frac{i}{\tau_{\alpha}(\boldsymbol{k})}\;.
\end{equation}
The wave of phonon populations is thus rewritten as
\begin{equation}
\Delta n_{\mu}^{\alpha}(\boldsymbol{k})=|I_{\mu}^{\alpha}(\boldsymbol{k})|e^{-t/\tau_{\alpha}(\boldsymbol{k})}\sin(\boldsymbol{k}\cdot\boldsymbol{x}-\overline{\omega}_{\alpha}(\boldsymbol{k})t+\phi)\;,\label{phonon_solution}
\end{equation}
where the phase shift $\phi$ arises from the imaginary part of $I$.
With this notation, the real part of the eigenvalue $\overline{\omega}$ is the oscillation frequency and the imaginary part $\tau$ is the relaxation time associated with the temperature wave.
We mention in passing that left and right eigenvectors of $B$ do not need to coincide and thus care must be taken in performing  algebraic manipulations.

So, the LBTE admits a basis set of solutions that correspond to temperature oscillations and, since the derivation doesn't rely on any specific assumption about the crystal in exam, temperature waves should exist in all crystals.
This observation may appear in contrast with the fact that second sound has been observed only in a handful of materials.
We speculate that this might be due to a number of practical difficulties.
First, relaxation times may often be too short to detect temperature waves before these are dissipated, and only in few circumstances (e.g. when normal processes dominate) the decay time might be long enough to allow for experimental observations on macroscopic time scales.
Another potential problem could arise if more than one mode is simultaneously excited during an experiment: in this case, the superposition of several modes could hide the original wave behavior.
We therefore remark that further investigations are needed to interpret second-sound experiments in the light of these considerations.

The long-wavelength limit of temperature waves is of relevance for macroscopic thermal conduction.
When $\boldsymbol{k}=0$, the matrix $B$ reduces to the scattering matrix and the eigenvalue problem is simplified to
\begin{equation}
\frac{1}{\mathcal{V}} \sum_{\mu'}\Omega_{\mu\mu'}I_{\mu'}^{\alpha}(\boldsymbol{k}=0) = i \omega_{\alpha}(\boldsymbol{k}=0)I_{\mu}^{\alpha}\;.
\end{equation}
Since $\Omega$ is real and symmetric \cite{relaxons}, the eigenvalues of $B$ at $\boldsymbol{k}=0$ are purely imaginary and determined by the relaxation time $\omega=\frac{i}{\tau_{\alpha}(\boldsymbol{k}=0)}$.
The eigenvalue problem thus reduces to the case of relaxons \cite{relaxons}.
Therefore, thermal transport in presence of macroscopic thermal gradients can be thought as originating from long-wavelength oscillations of temperature and phonon populations.

%%%%%%%%%%%%%%%%%%%%%%%%%%%%%%%%%%%%%%%%%%%%%%%%%%%%%%%%%%%%%%%

\section{Relation with previous literature}

We now compare this model of second sound with the studies of Guyer and Krumhansl (GK) \cite{guyer:ss} and Hardy \cite{hardy-ss} (which paved the way to many studies of second sound, this one included). 
We briefly recall here that GK's approach is based on the representation of the phonon population in terms of the eigenvectors of the momentum conserving part of the scattering matrix. 
Their work takes advantage of the fact that $D+1$ of these eigenvectors are analytically known, $D$ being the dimensionality of the system.
Of these eigenvectors, $D$ are associated with the drifting distribution $\bar{n}^{\text{drift}}_{\mu}=\frac{1}{\exp((\hbar\omega_{\mu}-\boldsymbol{q}\cdot\boldsymbol{V}^{\text{drift}})/k_{B}T)-1}$, with $\boldsymbol{V}^{\text{drift}}$ the drift velocity (a Langrange multiplier that conserves momentum); the remaining eigenvector is associated with the Bose--Einstein distribution (more on this later).
Hardy instead studied second sound by first writing the BTE in the complete basis of eigenvectors of the scattering matrix and then Fourier transforming the resulting equation.
We refer to their works for more details.

Some improvements are aimed at reducing the number of hypothesis required to model second sound.
In contrast with GK's work, the present work (a) avoids the single relaxation-time approximation and uses the complete scattering matrix, (b) employs the complete dispersion relation rather than the Debye approximation and (c) is not restricted to isotropic crystals. 
Hardy limited the discussion to an isotropic crystal, whereas the present discussion applies to any crystal symmetry.

There are also qualitative differences in the results, for example in the number of second sound modes found.
Direct diagonalization approaches, such as the present one, provide a complete basis set for the solution of the BTE. 
As a result, we have shown the existence - in the thermodynamic limit - of an infinite number of modes. 
In contrast, GK only found $D$ modes as a consequence of their choice of an incomplete basis set, driven by their lack of knowledge of the complete eigenvalue spectrum.
The physical implication of this basis set truncation has been noted by Hardy \cite{hardy-ss}: within GK's approach, second sound can only exist when momentum-conserving normal processes dominate over momentum-dissipating Umklapp events.
However, this conclusion is limited by the fact that the basis set is incomplete and that the basis functions are already associated with the conservation of momentum. 
%Therefore, GK concluded that the condition of existence of second sound is a slow decay of the momentum flux.
By using a complete basis set, Hardy managed to construct second sound modes that do not require an approximate momentum conservation: the more general condition for observing second sound is the slow decay of the energy flux, GK's condition being a special case where both energy and momentum fluxes decay slowly.
For this reason, Hardy was able to derive 3$D$ second sound modes without excluding the existence of other solutions: indeed, the present work finds an infinite number of modes.
%We note in passing how it is desirable to provide an explanation of second sound that doesn't rely on the distinction between normal and Umklapp processes, since their definition is not unique and depends on the choice of the unit cell.
We note in passing that the definition of normal and Umklapp processes is not unique, since it depends on the choice of the unit cell; it is therefore desirable to provide explanations of second sound that do not rely on a distinction between the two.

%Moreover, Hardy proceeded by looking for approximate solutions of the Fourier transform of the BTE, whereas we recognized and solve exactly an eigenvalue problem.

As a last difference, it was found in Ref. \cite{relaxons} that the function $\theta^0_\mu = \sqrt{\frac{\bar{n}_\mu(\bar{n}_\mu+1)}{Ck_BT^2}}\hbar \omega_\mu$ derived from the Bose--Einstein distribution is not an eigenvector of the scattering matrix with zero eigenvalue (see Ref. \cite{relaxons} for more details).
In contrast, both works from GK and Hardy mistakenly treated $\theta^0_\mu$ as an eigenvector, therefore their works should be revised taking into account that $\sum_{\mu'} \tilde{\Omega}_{\mu\mu'} \theta^0_{\mu'} \ne 0$.
In particular, GK's results should be revised after Eq. 20 of Ref. \cite{guyer:ss}, and Hardy's analysis should be reexamined from Eq. 5.3 of Ref. \cite{hardy-ss}.

To conclude this section, it's worth commenting on the limits of applicability of the modeling presented in this work.
All the results have been derived as exact solutions of the LBTE, which is assumed to hold throughout our manuscript. 
Therefore, we expect our results to be correct in systems where conduction is limited by scattering events; instead, ballistic regimes typically fall outside the domain of applicability of the (L)BTE and are more conveniently described using other formalisms such as Green's function techniques.
Moreover, the semiclassical approximation at the base of the (L)BTE requires that the wavelength of the perturbation ($\lambda = \frac{2\pi}{|\boldsymbol{k}|}$) to be much larger than the spread of the phonon wavepacket that is used to derive the (L)BTE \cite{ashcroft}.
As a result, the (L)BTE only holds for $\lambda$ much larger than the lattice constant: the transport waves obtained here lose their meaning as one moves closer to the Brillouin zone edge. In fact, one can note that the dispersion relation $\omega_\alpha(\boldsymbol{k})$ doesn't obey the periodicity of the Brillouin zone. Therefore, we expect the transport wave solutions to be suitable for describing long-wavelength excitations, whereas an alternative microscopic model is needed to accurately study such modes at small wavelengths.

%%%%%%%%%%%%%%%%%%%%%%%%%%%%%%%%%%%%%%%%%%%%%%%%%%%%%%%%%%%%%%%

\section{Computational case study: graphene}

As a practical example, we compute the dispersion relations of temperature waves in graphene.
Lattice harmonic and anharmonic properties are computed using density-functional perturbation theory \cite{baroni:rev,giannozzi:prb43,debe:prl,paulatto,lazzeri:anharmonic,baroni:prl-dfpt,gonze:2np1} as implemented in the Quantum ESPRESSO distribution \cite{qe}.
The scattering matrix is built with three-phonon and phonon-isotope interactions (at natural carbon abundances) over a 128$\times$128$\times$1 grid of points in the Brillouin zone.
The matrix $B$ is fully diagonalized numerically using subroutines of the ScaLAPACK library \cite{scalapack}, and calculations are managed using the AiiDA materials' informatics platform \cite{AiiDA}.
For all the remaining details, we refer to Ref. \cite{relaxons}, where the same computational parameters were used.

\begin{figure}
\centering
\includegraphics{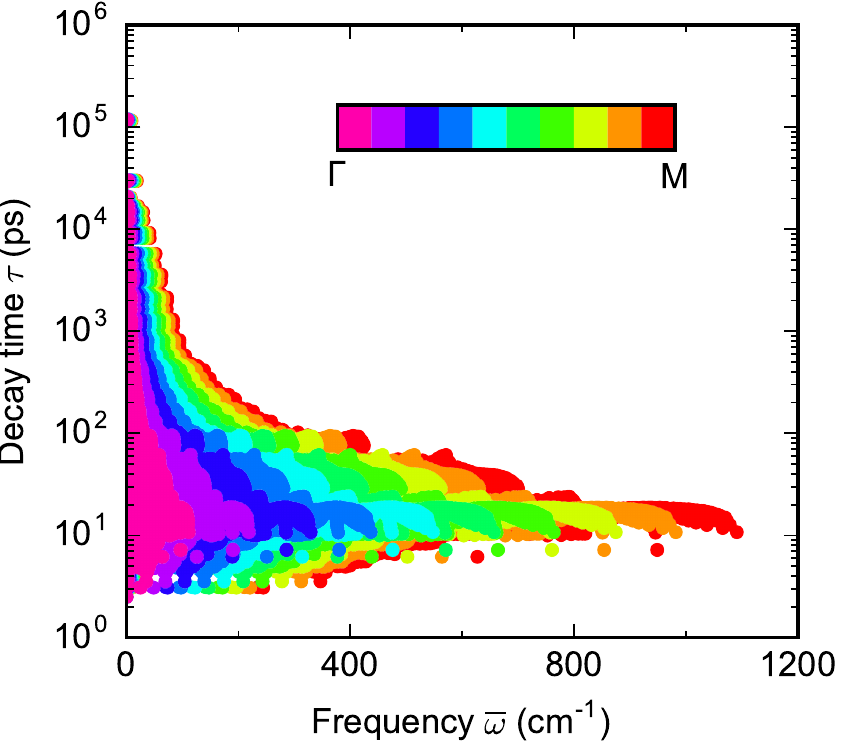}
\caption{
Eigenvalues $\omega = \bar{\omega} - \frac{i}{\tau}$ of graphene as a function of the oscillation frequency $\bar{\omega}$ (horizontal axis) and the relaxation time $\tau$ (vertical axis); colors identify different k-points in the Brillouin zone lying between $\Gamma$ and the edge M.
The allowed oscillation frequencies and decay times are both bounded.
Frequencies have a strong variation with the wavevector: in the limit of $\boldsymbol{k}=0$ all frequencies must vanish and the range of allowed frequencies increases with the wavevector.
}
\label{fig1}
\end{figure}

We first show in Fig. \ref{fig1} the eigenvalue spectrum of a few wavevectors between the Brillouin zone center $\Gamma$ and the edge point M.
In particular, we plot the decay time $\tau_{\alpha}$ of the eigenvalues as a function of the oscillation frequency $\bar{\omega}_{\alpha}$ for all temperature waves computed ($\approx10^{5}$ in the calculations).

The majority of temperature waves have relaxation times of about 10 ps: it's at these values that oscillating frequencies cover the broadest range, with $\bar{\omega}$ varying from 0 to about 800 cm$^{-1}$.
For simplicity, the negative frequency part of the spectrum is not shown, since by symmetry both $\bar{\omega}+\frac{i}{\tau}$ and $-\bar{\omega}+\frac{i}{\tau}$ are eigenvalues: this is simply a consequence of the Fourier transform (both signs are needed when building a sinusoidal solution).

The smallest decay times are set by the fastest rates of phonon scatterings: since the most frequent events occur every few picoseconds, there can not exist a temperature wave with a faster decay.
At the top of the scale in Fig. \ref{fig1}, the microsecond time scale is the same of the longest relaxation times of relaxons, i.e. of the modes at the Brillouin zone center.
This time scale can be much longer than that of phonon scattering, because heat flux may not be thermalized by a single scattering event \cite{relaxons} and thus heat flux can propagate over several scattering events before it's eventually dissipated.
We also note that the longest lived temperature waves are characterized by the smallest oscillation frequencies, and that they decay at a faster rate as frequency is increased.

The eigenvalue spectrum seems to large extent continuous; however, there are a few isolated features at the bottom and top of the picture.
Unfortunately, computational costs prevent a systematic convergence study of these isolated modes.
Only in the $\boldsymbol{k}=0$ limit the purely imaginary matrix is simpler to diagonalize and in Ref. \cite{relaxons} we could inspect convergence, concluding that the longest lived relaxons form a discrete spectrum.
Therefore, we speculate that the eigenvalue spectrum of temperature waves consists of a continuum and of a discrete set.

The spectra of $B$ at various $\boldsymbol{k}$ points are similar.
The most important difference is that, as $\boldsymbol{k}$ diminishes in magnitude, the range covered by $\bar{\omega}$ narrows until it reduces to 0 in the limit $\boldsymbol{k}=0$.
The spread in the relaxation times instead does not display significant changes.

%%%%%%%%%%%%%%%%%%%%%%%%%%%%%%%%%%%%%%%%%%%%%%%%%%%%%%%%%%%%%%%%%%%%%%%%%%%%%%%%%%

\begin{figure}
\centering
\includegraphics{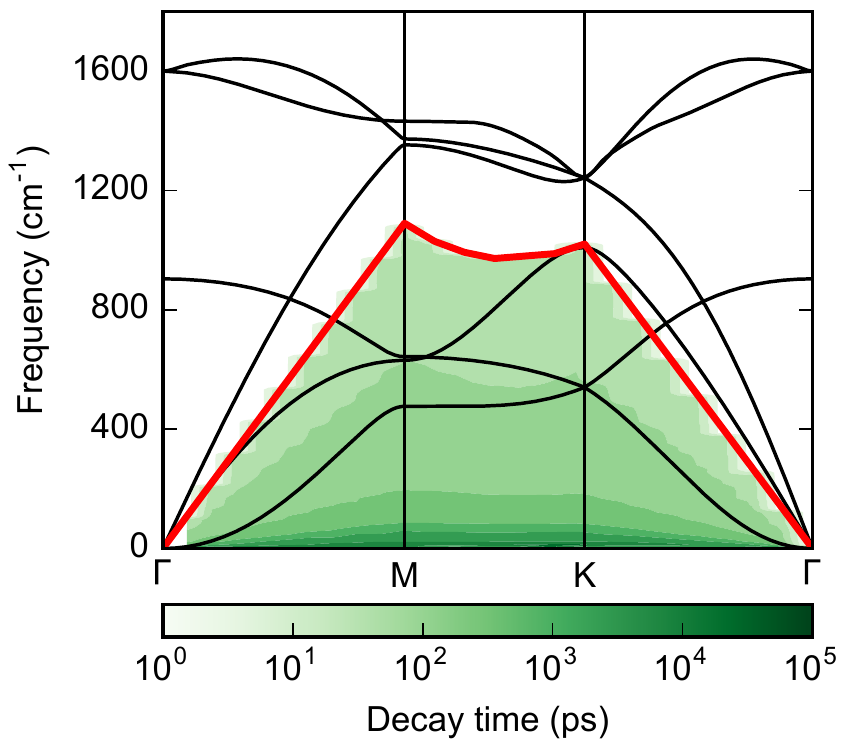}
\caption{
Continuum spectrum of dispersion relations for temperature waves in graphene at room temperature (shades of green); the red line marks the largest value for the allowed oscillation frequency, and all frequencies lying below are valid solutions. Phonon dispersions at zero temperature are also plotted (black lines). The different shades of green indicate the decay times of the temperature waves, showing how such modes can be longer lived, especially in the low-frequency region.}
\label{fig2}
\end{figure}

Last, in Fig. \ref{fig2} we plot the dispersion relations of the temperature waves over a high-symmetry path in the Brillouin zone; for comparison, the phonon dispersions are reported as black lines.
The red line represents the largest oscillation frequency for a given wavevector $\boldsymbol{k}$, and is such that for all frequencies below this line a corresponding temperature wave exists.
The largest oscillating frequency is roughly linear with the wavevector, with a slope approximately isotropic and similar to that of the transverse acoustic mode at $\Gamma$.
The linearity is reasonable in view that the real part of $B$ is linear in $\boldsymbol{k}$.
In the long wavelength limit, the temperature wave frequency drops to zero and only the imaginary part of the spectrum survives.

The colored areas indicate the values of the largest relaxation times available at a given frequency.
We can now see over the entire path in the Brillouin zone that the largest relaxation times are, as hinted in Fig. \ref{fig1}, roughly monotonic with the frequency: the slowest decaying modes all have small frequencies.
We thus speculate that experimental observation could be more likely for these low lying modes, since these can propagate undamped for the longest time.
The step-like features in the colored areas are numerical artifacts, due to the fact that only 25 wavevectors are computed along the high-symmetry path.

%{\color{blue}OLD: Finally, we remark that the temperature waves discussed here do not obey the periodicity of the Brillouin zone, due to a shortcoming of the BTE and LBTE.
%In fact, in Eq. \ref{phonon_bte} it was assumed that the phonon excitation number can be defined at any continuum point $\boldsymbol{x}$ in the real space.
%However, lattice vibrations can hardly be defined over distances that are smaller than the interatomic distance: the physically relevant wavevectors are only those belonging to the first Brillouin zone.
%Due to the continuum formulation, temperature waves do not reflect the folding scheme of the Brillouin zone and, unlike phonons, the dispersion of large $\boldsymbol{k}$ vectors does not fold into the first Brillouin zone.
%}
%{\color{red}NEW:

%%%%%%%%%%%%%%%%%%%%%%%%%%%%%%%%%%%%%%%%%%%%%%%%%%%%%%%%%%%%%%%%%%%%%%%%%%%%%%%%%

\section{Conclusions}

In conclusions, we have shown in all generality how to construct transport waves from the linearized Boltzmann transport equation.
By looking for wave-like solutions of the populations in the LBTE, we reduced the equation to a complex eigenvalue problem where each eigensolution corresponds to a 
transport wave.
These transport waves have well-defined wave-vectors, oscillation frequencies, relaxation times and dispersion relations, thus behaving as crystal excitations.
For the case of thermal transport, these excitations correspond to heat or temperature waves and in the long-wavelength limit give rise to relaxons, i.e.
the heat carriers for steady-state thermal transport.
We studied in detail graphene at room temperature, showing that an entire basis-set of wave solutions exists, with modes that are characterized by relaxation times that can reach microseconds.
Since the entire formalism is not tied to any assumption besides the validity of the LBTE, we conclude that temperature waves, and more generally,
transport waves are excitations present in principle in any crystal, and whose observation would rely on a combination of sufficiently long relaxation times
and frequency-resolved experimental techniques.

We gratefully acknowledge the Swiss National Science Foundation under Project ID 200021\_143636, the National Centre of Competence in Research MARVEL, the Max Planck-EPFL Center for Molecular Nanoscience and Technology, and the Swiss National Supercomputing Center CSCS under Project ID s580.

\section*{Appendix}
As we discussed in Ref. \cite{relaxons}, the matrix $\Omega$ appearing in Eq. 2 is not symmetric, i.e. $\Omega_{\mu\mu'} \neq \Omega_{\mu'\mu}$.
Nevertheless, a simple scaling of the LBTE is sufficient to bring it in a symmetrized form.
To this aim, let us define:
\begin{gather}
\tilde{\Omega}_{\mu\mu'} =  \Omega_{\mu\mu'}   \sqrt{ \frac{\bar{n}_{\mu'} (\bar{n}_{\mu'}+1)}{\bar{n}_{\mu} (\bar{n}_{\mu}+1) }} \;, \text{ and}
\label{n_symmetrisation} \\
\Delta \tilde{n}_{\mu} =  (\bar{n}_{\mu} (\bar{n}_{\mu}+1) )^{-\frac{1}{2}} \Delta n_{\mu}\;.
\label{n2_symm}
\end{gather}
With this variable change, the BTE becomes:
\begin{equation}
\frac{\partial\Delta \tilde{n}_{\mu}(\boldsymbol{x},t)}{\partial t}+\boldsymbol{v}_{\mu}\cdot\nabla(\Delta \tilde{n}_{\mu}(\boldsymbol{x},t))=-\frac{1}{\mathcal{V}}\sum_{\mu'}\tilde{\Omega}_{\mu\mu'}\Delta \tilde{n}_{\mu'}(\boldsymbol{x},t)\;.
%\label{phonon_bte2}
\end{equation}
The structure of the equation remains the same, but now the matrix $\tilde{\Omega}$ can be shown to be symmetric.
The discussion of temperature waves as shown in the main text applies identically to this modified equation.
One can thus define a symmetric matrix $\tilde{B}_{\mu\mu'} = \boldsymbol{k}\cdot \boldsymbol{v}_\mu \delta_{\mu\mu'} - \frac{i}{\mathcal{V}} \tilde{\Omega}_{\mu\mu'}$ and solve an eigenvalue problem, yielding eigenvalues $\omega$ and eigenvectors $\tilde{I}$.
In addition, the eigenvalues are left unchanged by the transformations.
The phonon population however must be scaled back via:
\begin{align}
  \Delta n_{\mu}^{\alpha}(\boldsymbol{k}) =& \sqrt{\bar{n}_\mu(\bar{n}_\mu+1)} |\tilde{I}_{\mu}^{\alpha}(\boldsymbol{k})|e^{-t/\tau_{\alpha}(\boldsymbol{k})}  \nonumber \\
  &\sin(\boldsymbol{k}\cdot\boldsymbol{x}-\overline{\omega}_{\alpha}(\boldsymbol{k})t+\phi)\;.
\end{align}

\bibliographystyle{apsrev4-1}
%\bibliography{scibib}

%

\end{document}